\documentclass[fleqn,twoside]{article}
\usepackage{espcrc2}
\usepackage{graphicx}
\usepackage[figuresright]{rotating}

\newcommand{\AmS}{{\protect\the\textfont2
  A\kern-.1667em\lower.5ex\hbox{M}\kern-.125emS}}

\title{Quantum first order phase transitions}

\author{Mucio A. Continentino  \address[IFUFF]{Instituto de F\'{\i}sica - Universidade Federal Fluminense\\
Av. Litor\^anea s/n,  Niter\'oi, 24210-340, RJ - Brazil}
        \thanks{Work partially supported by
the Brazilian Agencies, FAPERJ and CNPq.}
  and Andr\'e S. Ferreira\addressmark[IFUFF]. }

\begin{document}

\begin{abstract}
The scaling theory of critical phenomena has been successfully
extended for classical first order transitions even though the
correlation length does not diverge in these transitions. In this
paper  we apply the scaling ideas to {\em quantum first
order transitions}. The usefulness of this approach is illustrated treating the problems of a
superconductor coupled to a gauge field  and of
a biquadratic Heisenberg chain, at zero temperature. In both cases there is a {\it latent heat}
associated with their discontinuous quantum  transitions.  We
discuss the effects of disorder and give a general criterion for
it's relevance in these transitions. \vspace{1pc}
\end{abstract}

% typeset front matter (including abstract)
\maketitle

\section{Introduction}

Scaling theories are invaluable tools in the theory of quantum critical phenomena \cite{mu,hertz}.  They yield
relations among the critical exponents governing the  behavior of relevant thermodynamic quantities
at very low temperatures. In the study of strongly correlated metals close to a quantum instability, they led to
the discovery of a new characteristic temperature which marks the onset of Fermi liquid behavior \cite{mu}. Here
we study the extension of scaling ideas to quantum first order  phase transitions~\cite{mucio}. Although there
is no diverging length in these transitions, this has proved to be very
useful~\cite{nienhuis,fisher,nbrs,sp} for temperature driven transitions and will turn out to be also the case
for discontinuous quantum transitions.

Let us consider the scaling form of the $T=0$ free energy density close to
the quantum phase transition,
\begin{equation}
\label{funda}
f \propto |g|^{2-\alpha}
\end{equation}
where $g$ measures the distance to the transition at $g=0$. The
exponent $\alpha$ is related to the correlation exponent $\nu$
through the quantum hyperscaling relation  $2-\alpha=\nu(d+z)$
\cite{mu} where $d$ is the dimension of the system and $z$ the
dynamic  critical exponent~\cite{mucio}. The total internal energy
close to the transition can be written as,
\begin{equation}
U(g=0^{\pm})=U(g=0) \pm A_{\pm}|g|^{2-\alpha}
\end{equation}
for $g \rightarrow 0^{\pm}$. Then the existence of a first order
phase transition at $T=0$ with a discontinuity in $dU/dg$ and a
{\it latent heat} implies the value $\alpha=1$ for this critical
exponent. If quantum hyperscaling applies, this leads to a
correlation length exponent $\nu=1/(d+z)$. This is the quantum
equivalent of the classical result $\nu=1/d$ for temperature
driven first order transitions~\cite{nienhuis,fisher,nbrs,sp}.
Associated with this value of the correlation length there is on
the disordered side of the phase diagram a new energy scale, $T^*
\propto |g|^{z/(d+z)}$.

The presence of a discontinuity in the order parameter and the assumption of no-decay of it's correlation function
imply $\beta=0$, as in the classical case~\cite{fisher} and $d+z-2+ \eta=0$, respectively. As for classical
transitions $\delta = \infty$ and for consistency with the scaling relations the order parameter susceptibility
seems to diverge with an exponent $\gamma=1$~\cite{fisher}.

If the quantum transition is driven, for example,  by pressure, $g
\propto (P-P_c)/P_c$ where $P_c$ is a critical pressure and a
finite {\it latent heat} means in this case a finite amount of
work, $W=A_{+}+A_{-}=P_c \Delta V$, to bring one phase into
another. Such finite {\it latent work } is associated with a
change in volume since the intensive variable, pressure in this
case, remains constant at the transition. In the case of a density
driven first order transition the chemical potential remains fixed
while the number of particles changes.

In the next sections we study two problems which present first order quantum transitions and confirm the results
obtained above on the basis of scaling arguments. These results are also important to clarify the meaning and
the range of
application of a scaling analysis in situations where criticality is in fact avoided.

\section{Superconductor coupled to a gauge field}

\begin{figure}[ptb]
%\begin{center}
\includegraphics[width=6.5cm]{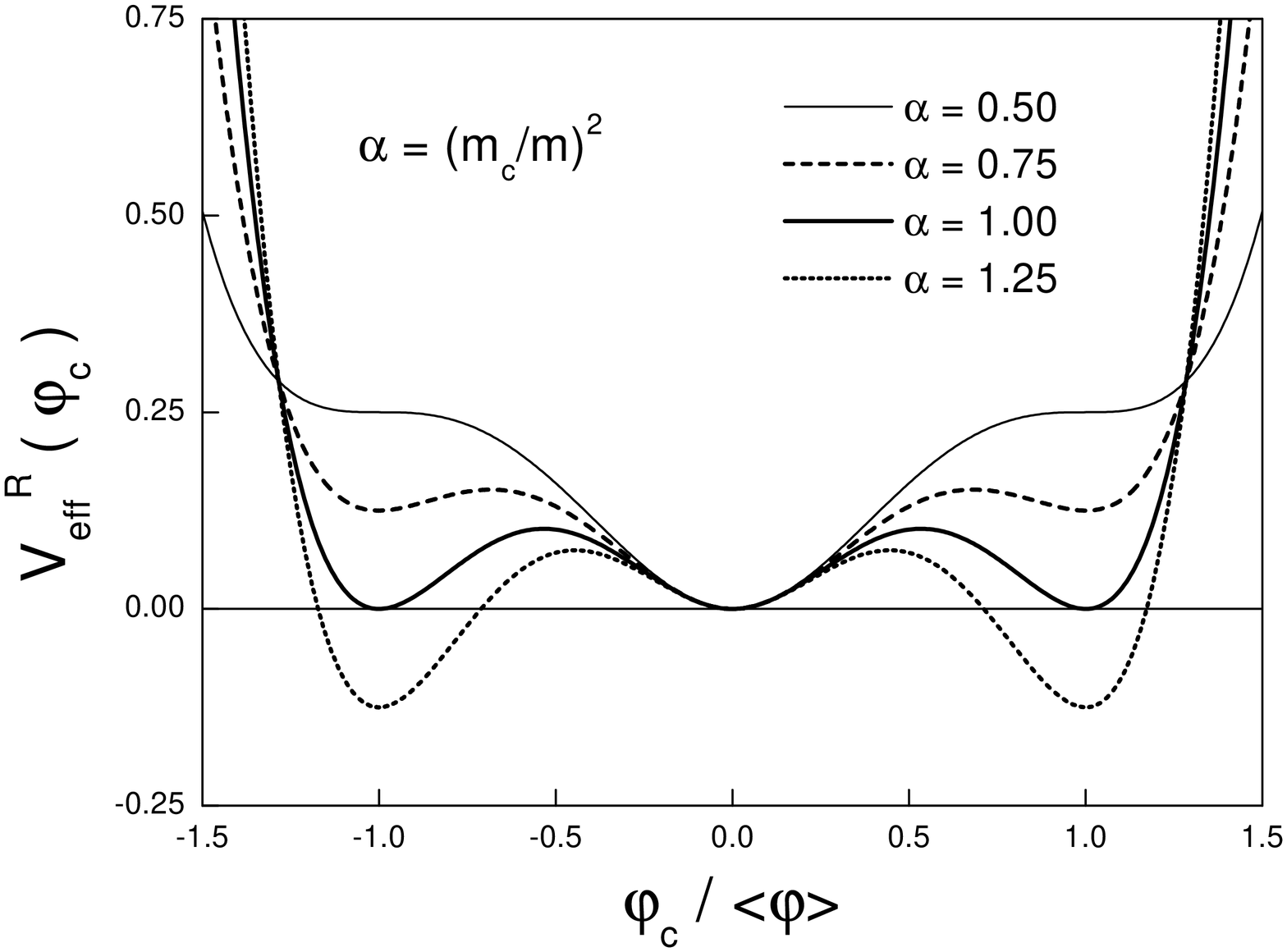}\caption{Zero temperature
first order transition in the Coleman-Weinberg
potential.} \label{fig1}
%\end{center}
\end{figure}

An interesting case of a quantum first order transition occurs in a superconductor coupled to the electromagnetic
field at $T=0$. The starting point to describe this transition is the Lagrangian density of charged particles
minimally coupled to the electromagnetic field. The Lagrangian density of the model in given by,
\begin{eqnarray}
L&=&-\frac{1}{4}(F_{\mu \nu })^{2}\!+\!\frac{1}{2}(\partial _{\mu
}\varphi _{1}\!+\!qA_{\mu }\varphi _{2})^{2}+\nonumber
\\
&&+\frac{1}{2}(\partial _{\mu }\varphi _{2} -qA_{\mu }\varphi
_{1})^{2}+\nonumber
\\
&& -\frac{1}{2}m^{2}(\varphi _{1}^{2}+\varphi
_{2}^{2})-\frac{\lambda }{4!} (\varphi _{1}^{2}+\varphi
_{2}^{2})^{2}
\end{eqnarray}
where the first term is the Lagrangian of the electromagnetic
field ($F_{\mu \nu }=\partial_{\mu}A_{\mu}-\partial_{\nu}A_{\nu}$)
and the complex scalar field $\varphi $ associated with the
superconducting state is given by, $\varphi
=\frac{1}{\sqrt{2}}(\varphi _{1}+i\varphi _{2})$, with $\varphi
_{1}$ and $\varphi _{2}$ real. At $T=0$ time enters as a new
direction and the indices $\mu,\nu$ run from $0$ to $d=3$. The
minimal coupling between these fields is through the electric
charge $q$ and we are working in $\hbar=c=1$ units. This is
essentially a quantum version of the Landau-Ginzburg free energy
of a superconductor in a magnetic field~\cite{colwein}. As we are
dealing with a Lorentz invariant case in which space and time
enter on equal footing in the Lagrangian density, we can identify
the dynamic critical exponent, $z=1$. For the chargeless problem
($q=0$) the Lagrangian above is associated with a quantum
superfluid-insulator transition  at $m^2=0$ as we discuss below.
Furthermore, we are interested here in the case of spatial
dimension $d=3$, such that, the effective dimension of the quantum
problem is $d_{eff}=d+z=4$. The zero temperature effective
potential associated with this Lagrangian density in the one-loop
approximation is given by~\cite{mucio,nogueira},
\begin{equation}
\label{v0}
V^0_{e\!f\!f}(\!\varphi_c\!)\!\!=\!\!\frac{m^2\varphi_c^2}{2}\!-\!
\frac{m^{2}\varphi _{c}^{4}}{4\langle\varphi\rangle ^{2}}
\!+\!\frac{3q^{4}\varphi _{c}^{4}}{64\pi ^{2}}[\ln (\frac{ \varphi
_{c}^{2}}{\langle\varphi\rangle ^{2}})\!-\!\frac{1}{2}] \label{vfinal}
\end{equation}
where $\varphi _{c}$ is the classical value of the field and
$\langle\varphi\rangle$ an extremum of the effective potential
defined such that $(dV_{eff}/d
\varphi_c)_{\varphi_c=\langle\varphi\rangle}=0$. When the mass
term vanishes the effective potential reduces to the
Coleman-Weinberg result \cite{colwein}. 
%Incidently, notice that
%the constant $\lambda$ has disappeared in the above equation for
%the effective potential due to the phenomenon of {\em dimensional
%transmutation} \cite{colwein,mucio}.  
In Fig.~\ref{fig1} we plot
the effective potential for different values of the mass $m^2$. At
a \textit{critical value} of the mass, $m_c^{2}$, given by,
\begin{equation}  \label{macrit}
m_c^{2} = \frac{3q^4}{32\pi^2}\langle\varphi\rangle^2
\end{equation}
there is a first order phase transition at zero temperature to a
new state of broken symmetry with $\varphi_c \ne 0$.

Let us examine how the energies of the different ground states
that exchange stability at the critical mass $m_{c}$ behave in the
neighborhood of the first order transition. For values of
$m>m_{c}$, the stable ground state, i.e. the minimum of the
effective potential, Eq.~(\ref{vfinal}), occurs when the order
parameter $\varphi_c =0$, such that, $V_{eff}(\varphi_c =0)=0$.
The value of the effective potential at the metastable minimum
$\varphi_{c}=\langle\varphi\rangle$ is given by,
\begin{equation}
V^0_{eff}(\langle\varphi\rangle)=%
\frac{1}{4}m^{2}\langle\varphi\rangle^{2}\left[
1-\frac{m_{c}^{2}}{m^{2}}\right] \label{first}
\end{equation}
Then at $m^{2}=m_{c}^{2}$ the two ground states at
${\varphi_{c}=0}$ and $\varphi _{c}=\langle\varphi\rangle$ are
degenerate and for $m^{2}<m_{c}^{2}$, the true ground state is at
$\varphi _{c}=\langle\varphi\rangle$. The effective potential at
$T=0$ represents the ground state energy \cite{jackiw} and close
to the critical mass $m_{c}$, we find,
$V_{eff}~\propto~|m^2-m_c^2|~\propto~|g|^{2-\alpha}$ which implies
that the critical exponent $\alpha =1$ and if hyperscaling applies
the correlation length exponent $\nu = 1/{(d+z)}$. The
\textit{latent heat} is given by
\[
L_{h}=(A_{+}+A_{-})=\frac{1}{4}m_{c}^{2}\langle\varphi\rangle^{2}
\]
where we used $A_{+}=0$. Notice the existence of a spinodal at $ (m_c/m)^2 =0.5$ which marks the limit of
metastability of the superconductor in the normal phase. On the other hand there is always a metastable minimum at
$\varphi_c=0$ in the superconducting phase.

\begin{figure}[hptb]
\begin{center}
\includegraphics[width=6.2cm]{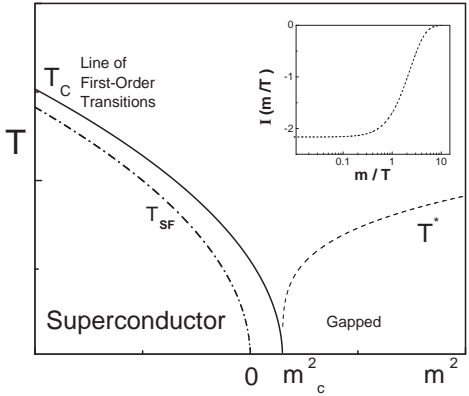}\caption{Finite
temperature phase diagram for a charged superfluid. \protect$T_{SF}$ is
the second order transition line for the neutral superfluid and \protect$T^*$
a new energy scale associated with the first order quantum transition. The inset shows the
function \protect$I(y)$ on Eq.~(\ref{VT}) for \protect$V_{eff}(T)$.}
\label{fig2}
\end{center}
\end{figure}

In order to treat the finite temperature case we note that for
quantum theories of Euclidean fields at finite temperatures, the
effective potential is equivalent to the thermodynamic free energy
\cite{jackiw}. The generalization of the effective potential to
$T\ne 0$ is done replacing frequency integrations in the
calculation of the effective potential by a sum over Matsubara
frequencies. The effective potential at finite $T$ ({$k_B=1$})
is given by,
\begin{eqnarray}
V_{eff}(T)&=&\frac{1}{4}m^2\langle\varphi\rangle^2 |g|\left\{1+ \right.\nonumber \\
&&+\!\!\left.\frac{2}{\pi^2m^2\langle\varphi\rangle^2}
\frac{T^{d+1}}{|g|}I\left(\frac{M(\varphi_c)}{T}\right)\right\}
\label{VT}
\end{eqnarray}
%\begin{eqnarray}
%V_{eff}(T)&=&V^0_{eff}(\varphi_c) +
%\frac{T^{d+1}}{2\pi^2}I\left(\frac{M(\varphi_c)}{T}\right)
%\label{VT}
%\end{eqnarray}
where $M^2(\varphi_c)=m^2 + q^2 \varphi_c^2$ and 
%and
%$V^0_{eff}(\varphi_c) $ is given in Eq. (\ref{v0}). 
\[
I_d(y)=\int_{0}^{\infty}dxx^{d-1}\ln[1-e^{-\sqrt{x^{2}+y^{2}}}].
\]
The function $I_3(y)=I(y)$ for three dimensions is plotted in
the inset of Fig.~\ref{fig2}. In the limit  $T\gg M$ and close to the critical
point, we have \cite{mucio},
\begin{eqnarray}
V_{eff}(\varphi_c,T) =-\frac{\pi
^{2}}{18}T^{4}-\frac{1}{8}m^{2}T^{2}+ \frac{1}{2}m_{T}^{2}\varphi
_{c}^{2} \nonumber \\
+\frac{\lambda}{24} \varphi_{c}^{4}+\frac{3q^{4}}{64\pi ^{2}}\varphi _{c}^{4}[\ln
(\frac{\varphi _{c}^{2}}{\Lambda^{2}})-\frac{25}{6}]
\end{eqnarray}
where we defined a renormalized temperature dependent mass,
\[ m_{T}^{2}=|m^{2}|(1-T^{2}/T_{MF}^{2}) \]
with ${T_{MF}^{2}\approx 12|m^{2}|/3q^{2}}$. Alternatively we can
write $m_T$ as
\begin{equation}
m_{T}^{2}=m^{2}+(q^{2}/4)T^{2}  \label{lc}
\end{equation}
and if we choose the arbitrary value of the quantity $\Lambda$ as the minimum,
 $\langle\varphi\rangle$, of the temperature dependent effective potential, we obtain
\begin{eqnarray}
V_{eff}(\varphi_c,T) =-\frac{\pi
^{2}}{18}T^{4}-\frac{1}{8}m^{2}T^{2}+ \frac{1}{2}m_{T}^{2}\varphi
_{c}^{2} \nonumber
\\
-\frac{m_{T}^{2}}{4\langle\varphi\rangle^{2}} \varphi
_{c}^{4}+\frac{3q^{4}}{64\pi ^{2}}\varphi _{c}^{4}[\ln
(\frac{\varphi _{c}^{2}}{\langle\varphi\rangle^{2}})-\frac{1}{2}]
\label{vvfinal}
\end{eqnarray}

Let us now discuss these results. First, notice from
Eq.~(\ref{lc}) that the line at which the temperature dependent
mass $m_T$ vanishes is given by,
\[
T_{MF}=\frac{2}{q}|m^2|^{1/2}
\]
If we consider the contribution of terms of $O(\lambda)$, this
temperature is in fact given by, ${T_{MF}^{2}=12|m^2 |/(4\lambda
+3q^{2})}$. This line has no special meaning  since,  on cooling the system a first order
transition occurs before it, as we show below
(see Fig.~\ref{fig2}). It is governed by the same (mean-field)
shift exponent, $\psi =z/(d+z-2)=1/2$, of the critical line of a
{\it neutral} superfluid given by $T_{SF}^{2}=12|m^2 |/(4
\lambda)$. Notice that in the $3d$-case, this  superfluid
insulator transition at zero temperature is described exactly by
the one-loop effective potential since $d+z=d_c= 4$ is the upper
critical dimension for this transition. This transition is the
interaction driven quantum superfluid-insulator transition studied
by Fisher et al.~\cite{matthew}. The insulating character of the
disordered phase is due to the presence of a gap for excitations,
$\Delta = |m^2|^{\nu z}=|m^2|^{1/2}$ since the correlation
exponent $\nu$ assumes it's mean-field value for $d=d_c$.

In the charged superfluid the actual transitions  are quite different and occur for
\begin{equation}
m_{T}^{2}=m^{2}+(q^{2}/4)T_{c}^{2}=m_{c}^{2}  \label{lc1}
\end{equation}
where $m_{c}^{2}$ is given by Eq.~(\ref{macrit}). The critical line of first order transitions is now given by,
\begin{equation}
T_{c}=\frac{2}{q}\sqrt{m_{c}^{2}-m^2 }  \label{lc2}
\end{equation}
and at the quantum critical point of the neutral superfluid, $m^2=0$, there is now a superconducting instability
at a finite critical temperature,
\[
T_{c}(m^2=0)=\sqrt{\frac{3}{8 \pi^2}} q\langle\varphi\rangle
\]
The physical origin of this phase transition is the energy gained by the
system with the expulsion of the electromagnetic field when the system
becomes superconducting.

%The generalization of Eq.~(\ref{first}) to finite $T$ ($k_B=1$)
%is given by,
%\begin{eqnarray}
%V_{eff}(T)&=&\frac{1}{4}m^{2}\langle\varphi\rangle^{2}|g|\bigg\{
%1+ \nonumber \\ & + &
%\frac{2}{\pi^{2}m^{2}\langle\varphi\rangle^{2}}\frac{T^{d+1}}{|g|}I\left(\frac{M(\varphi_c)}{T}\right)\bigg\}
%\label{VT1}
%\end{eqnarray}
From the temperature dependent effective potential of Eq.~(\ref{VT}) and the plot of the function $I(y)$ in
the inset of
Fig.~\ref{fig2} we conclude that there are {\em two relevant scales} for
the present problem in the disordered phase ($\varphi_c=0$, $m^2
> m_c^2$). For $m/T \gg 1$, the thermal contribution to the
effective potential vanishes exponentially as can be easily
checked.  For $m/T<0.12$, which
corresponds to high temperatures $I(y)$ saturates, $I(y  < 0.12
)\approx - 2.16$. In this case the effective potential,
\begin{eqnarray*}
V_{eff}(T) \approx \frac{1}{4}m^{2}\langle\varphi\rangle^{2}|g|\bigg\{
1-\frac{4.32}{\pi^{2}m^{2}\langle\varphi\rangle^{2}}\frac{T^{d+1}}{|g|}\bigg\}
\end{eqnarray*}
which can be cast in the scaling form,
$$
V_{eff}(T) \propto |g|^{2-\alpha} F \left[ \frac{T}{T^*} \right]
$$
with the critical exponent $\alpha=1$ and the characteristic temperature,
$$T^* \propto |g|^{\frac{z}{d+z}} = |g|^{\frac{1}{d+1}}=|g|^{\frac{1}{4}}.$$
This is similar to that of continuous quantum phase transitions, where
$T^* \propto |g|^{\nu z}$~\cite{mucio} but with $\nu=1/(d+z)$
confirming the expectations of our previous discussion. Notice that in the present problem,
the mass $m$ (or $m^2$), the control parameter itself, provides the natural cut-off for breakdown of
scaling along the temperature axis. The two characteristic energies $T^*$ and $m^2$, the scaling temperature
and the cut-off scale are general features expected to play a role near quantum
discontinuous transitions.

The quantum mechanical problem of two coexisting phases at $m_c$, the superconductor and the insulator,  can be
described by a double wave function $\psi = a \psi_1 + b \psi_2$. In the probability density $|\psi|^2 a^2|\psi_1|^2 + b^2|\psi_2|^2 + ab(\psi_1 \psi_2^* + \psi_2 \psi_1^*)$, the coefficients $a^2$ and $b^2$ are the
relative proportions of each phase. The interference term may have experimental significance as one of the phases
has macroscopic coherence. Even if the overlap between the wave functions vanishes in the
thermodynamic limit, at the first order transition it may give rise to finite corrections
as the system is made up of finite domains due to the avoided criticality.

\section{The biquadratic chain}

The transition investigated above is a special case of quantum first order
transitions referred as fluctuation induced first order
transitions. From the scaling analysis we expect however that
the result $\nu = 1/(d+z)$ holds  generally for transitions
with a latent heat.
As an example that this is the case we
investigate the biquadratic spin-1 chain~\cite{batchel},
\begin{equation}
H = - \sum_i \epsilon_i (\vec{S}_i\cdot\vec{S}_{i+1})^2
\end{equation}
with
\begin{equation}
\epsilon_i = \left\{ \begin{array}{ll}
            1  & \mbox{if $i$ is odd} \\
           \lambda & \mbox{if $i$ is even}
          \end{array}
        \right.
\end{equation}
At $\lambda=1$ there is a zero temperature first order phase
transition where two spontaneously dimerized ground states
exchange stability~\cite{batchel}. The ground state energy can be written as
\begin{equation}
E_g(\lambda)-E_g(\lambda=1)= A_{\pm}|1-\lambda|
\end{equation}
consistent with $\alpha=1$ and the latent heat ${L=-(A_++A_-)}$
can be exactly obtained~\cite{batchel}. Furthermore in this case
the correlation length exponent has been directly obtained from finite
lattice calculations~\cite{soly}. The numerical value,  $\nu
\approx 0.5$ agrees with the expected value $\nu=1/(d+z)=1/2$
since for this transition $z=1$~\cite{soly}.

\section{Effects of Disorder}

The effects of disorder on classical first order transitions
have been extensively studied~\cite{cardy}.
Here we
must distinguish weak or fluctuation-induced first order
transitions from strong first order transitions which map into the
random field problem~\cite{cardy} since disorder couples to the
order parameter.
In the latter case a criterion for the role of disorder
based on domain wall
arguments can be easily generalized to quantum systems. In
analogy with the random field problem~\cite{biarf}, we define a generalized
stiffness $J$ associated with the (continuous) symmetry-broken
phase~\cite{mucio}, which scales as $J^{\prime}= b^{d+z-2}J$
close to the strong coupling attractor of this phase.
In this
equation $b$ is the scaling factor and $d$ and $z$ are
respectively the dimension and the dynamical critical exponent
discussed earlier.
At this fixed point the random field scales as $h^{\prime}=b^{d/2}h$~\cite{biarf},
and their ratio
\begin{equation}
\left( \frac{h}{J} \right)^{\prime}=b^{\frac{4-(d+2z)}{2}}\left( \frac{h}{J} \right)
\end{equation}
If $d+ 2z > 4$ the fixed point at $(h/J)=0$ is stable and a
critical amount of disorder is required to destroy the ordered
phase.  As concerns the first order transition this implies that
the coexistence of phases is possible at least for sufficiently
weak disorder. In the opposite case, i.e., for $d+ 2z < 4$
disorder destroys the first order character of the transition
since there can be no coexistence and one phase grows at the
expenses of the other. The case $d+ 2z = 4$ is marginal and
requires specific calculations. For the biquadratic chain
discussed above $d+ 2z =1+2=3 < 4$ and any amount of disorder
drives this system to a random singlet phase associated with an
infinite disorder fixed point as has been shown using a
perturbative renormalization group calculation~\cite{bia}.

In the case of fluctuation induced quantum first order transitions
we find no general criterion as for the standard case. For
the problem treated here of a superconductor coupled to a gauge
field, Boyanovsky and Cardy~\cite{boya} have shown that to order
$4 - \epsilon$, at least for weak disorder, this transition
remains first order.

\section{Conclusions}

We have investigated quantum first order transitions using scaling ideas and
looking at two
specific cases. In both cases there is a discontinuity in the
first derivative of the ground state energy equivalent to a latent
heat associated with the transition. Therefore, as in classical
transitions, we have $\alpha=1$ which allows the definition of the
correlation length exponent $\nu = 1/(d+z)$ for the quantum case.
The consideration of the problem of the superconductor coupled to a gauge field
has been important to clarify the meaning of a scaling approach in a system where criticality
is avoided.
We have studied the effects of disorder in these transitions in the case
this couples to fluctuations of the order parameter. A simple
criterion to determine if the first order nature of the quantum
transition is modified by disorder is discussed.


\begin{thebibliography}{9}

\bibitem{mu} M. A.
Con\-ti\-nen\-ti\-no, G. Ja\-pias\-su and A. Tro\-per, Phys. Rev. \textbf{B39
\/},  9734 (1989); M.A. Con\-ti\-nen\-ti\-no, Phys.
Rev. \textbf{B47 \/}, 11587 (1993).

\bibitem{hertz} J. Hertz, Phys. Rev. {\bf B14}, 1165 (1976).

\bibitem{mucio} M. A. Continentino, \emph{Quantum Scaling in Many Body
Systems}, World Scientific, Singapore, 2001.


\bibitem{nienhuis}
    B. Nienhuis and N. Nauenberg, Phys. Rev. Lett.
    {\bf 35}, 477 (1975).

\bibitem{fisher}
    M. E. Fisher and A. N. Berker, Phys. Rev. B
    {\bf 26}, 2507 (1982).

\bibitem{nbrs}
    B. Nienhuis, A. N. Berker, E. K. Riedel, and
    M. Schick, Phys. Rev. Lett. {\bf 43}, 737 (1979).

\bibitem{sp}
    J. S\'olyom and P. Pfeuty, Phys. Rev. B {\bf 24},
    218 (1981); L. Turban and F. Igloi, Phys. Rev. B {\bf 66}, 014440 (2002).


\bibitem{colwein} S. Coleman and E. Weinberg, Phys. Rev. {\bf D7}, 1888 (1973); B. I. Halperin, T. C. Lubensky and S.
Ma, Phys. Rev. Lett. {\bf 32}, 292 (1974).

\bibitem{nogueira} A.P.C. Malbouisson, F. S. Nogueira and N.F. Svaiter, Mod. Phys. Letts. {\bf A11}, 749 (1996).

\bibitem{jackiw} R. Jackiw, Phys. Rev. {\bf D9}, 1686 (1973).

\bibitem{matthew} M. P. A. Fisher,  P.B. Weichman,  G. Grinstein and  D. S. Fisher,
{\it Phys. Rev.  \/}{\bf B40}, 546 (1989).

\bibitem{batchel} M. Barber and M T. Batchelor, Phys. Rev. {\bf B40}, 4621 (1989).

\bibitem{soly} J. S\'olyom, Phys. Rev. {\bf B36}, 8642 (1987).

\bibitem{cardy} see J. L. Cardy, Physica {\bf A263}, 215 (1999) and references therein.

\bibitem{biarf} B. Boechat and M.A. Con\-ti\-nen\-ti\-no, J. Phys: Condensed
Matter, {\bf 2 \/}, 5277 (1990).


\bibitem{bia} B. Boechat, A. Saguia e M.A. Continentino, Sol.
St. Comm. {\bf 98}, 411 (1996).

\bibitem{boya} D. Boyanovsky and J. L. Cardy, Phys. Rev. {\bf B25}, 7058 (1982).

\end{thebibliography}
\end{document}